\documentclass{revtex4}

\usepackage{graphicx}
\usepackage{bm}
\usepackage{amssymb}
\usepackage{amsmath}
\usepackage{dcolumn}
\usepackage[latin1]{inputenc}
\usepackage[english]{babel}
\begin{document}

\title{Envelope Theory for Systems with Different Particles}

\author{Claude \surname{Semay}}
\email[E-mail: ]{claude.semay@umons.ac.be}
\author{Lorenzo \surname{Cimino}}
\email[E-mail: ]{lorenzo.cimino@student.umons.ac.be}
\author{Cintia \surname{Willemyns}}
\email[E-mail: ]{cintia.willemyns@umons.ac.be}
\affiliation{Service de Physique Nucl\'{e}aire et Subnucl\'{e}aire,
Universit\'{e} de Mons,
UMONS Research Institute for Complex Systems,
Place du Parc 20, 7000 Mons, Belgium}
\date{\today}

\begin{abstract}
\textbf{Abstract} The eigensolutions of many-body quantum systems are always difficult to compute. The envelope theory is a method to easily obtain approximate, but reliable, solutions in the case of identical particles. It is extended here to treat systems with different particles (bosons or fermions). The accuracy is tested for several systems composed of identical particles plus a different one.  
\end{abstract}

\maketitle

\section{Introduction}
\label{sec:intro}

The determination with a high accuracy of eigenvalues and eigenvectors of many-body quantum systems is always a hard problem requiring generally heavy computations. Several methods have been developed to tackle this issue. Among the most efficient, one can note the expansions in large oscillator basis \cite{silv96} or in correlated Gaussian basis \cite{horn14}, the Lagrange-mesh method \cite{timo17}, etc.  

The envelope theory (ET) \cite{hall80,hall83,hall04}, also known as the auxiliary field method, is a simple technique to compute approximate eigenvalues and eigenvectors for $N$-body Hamiltonians. The method has been extended to treat arbitrary kinematics in $D$ dimensions with various potentials, but only for identical particles \cite{sema13,sema17,sema18b,sema19}. Numerical approximation can always be easily computed. Moreover, in the most favourable cases, analytical lower or upper bounds can be obtained. Besides two-body interactions \cite{sema13,sema17}, a special type of many-body forces \cite{sema18b} can be handled. The accuracy of the method has been checked for $D=1$ \cite{sema19} and $D=3$ \cite{sema15a,sema15b} for various potentials. In particular, two-body interactions with a repulsive short range part are considered for $D=1$ with the Calogero model \cite{sema19} and for $D=3$ with confined bosons \cite{sema15a}. Many tests have been performed with the ground state, but the spectra can be calculated as well. An example is given in \cite{sema15a} for a three-quark system for which an accuracy of a few percent has been reached for the 16 lowest levels. Not only approximations of the eigenvalues can be obtained with the ET but also approximations of the eigenvectors. In \cite{sema15a}, mean values for the interparticle distance and values of the pair correlation function at the origin are computed for various systems with a reasonable accuracy.

The key element of the ET is the fact that the complete solution of a $N$-body harmonic oscillator Hamiltonian with one-body and two-body forces, says $H_\textrm{ho}$, can be obtained by the diagonalisation of a matrix of order $(N-1)$ \cite{silv10}. This diagonalisation can be performed analytically in many situations. If $H$ is the general $N$-body Hamiltonian under study, the idea is to build an auxiliary Hamiltonian, $\tilde H = H_\textrm{ho} + B$, where $B$ is a function of the masses and coupling constants of $H_\textrm{ho}$. The form of $B$ is unequivocally determined from the structure of $H$. The eigenvalues of $\tilde H$ are simply the eigenvalues of $H_\textrm{ho}$ plus the value of the function $B$. By using an extremisation procedure for the the masses and coupling constants of $\tilde H$, its eigenvalues can be rendered very close to those of $H$. The procedure to build $\tilde H$ and compute the approximate solutions of $H$ is described in the following sections. The detailed justification of the method is too long to be reproduced here but it is given in \cite{sema08}, where the other name of the envelope theory, the auxiliary field method, is also justified. Moreover, the approximate eigenvalues obtained with the ET are upper or lower bounds \cite{hall80,hall83} of the eigenvalues of $H$ in many cases.

In this paper, it is shown that the ET can be extended to obtain relevant information about $N$-body systems with different particles. If the eigenvalues of the Hamiltonian $\tilde H$, which is also a harmonic oscillator Hamiltonian, can be analytically obtained, the computation of approximate eigenvalues of the Hamiltonian $H$ under study is reduced to the determination of an extremum or a saddle point of a multivariate function. If $\tilde H$ cannot be exactly solved, the calculations must then include diagonalisation procedures of a matrix of order $(N-1)$. In Sect.~\ref{sec:gen}, the ET is developed for the general case of a many-body system containing an arbitrary number of different particles. The procedure to carry out calculations in a practical manner is explained in Sect.~\ref{sec:pract}, where an improvement of the method is also presented. In Sect.~\ref{sec:np1}, various systems with only one particle different from the other ones are studied to test the efficiency of the ET. Some concluding remarks and perspectives are given in Sect.~\ref{sec:conclu}.

\section{General theory}
\label{sec:gen}

In the $D$-dimensional space, the Hamiltonian $H$ is written as
\begin{equation}
\label{HN}
H = \sum_{i=1}^N T_i(p_i) + \sum_{i=1}^N U_i(s_i) + \sum_{i<j=2}^N V_{ij}(r_{ij}).
\end{equation}
The momentum of the $i$th particle, $\bm p_i$, is the conjugate variable of the position $\bm r_i$. A variable $\bm s_i=\bm r_i - \bm R$ is defined with respect to the centre of mass position $\bm R$ (see Appendix~\ref{sec:cm}), and $\bm r_{ij}=\bm r_i - \bm r_j$ as usual. Moreover, $p_i=|\bm p_i|$, $s_i=|\bm s_i|$ and $r_{ij}=|\bm r_{ij}|$. The kinetic part is arbitrary but some constraints exist for the possible forms of $T_i$: it must be a positive quantity which is an increasing function of only the modulus of the momentum, with some degree of differentiability \cite{sema18a}. One-body potentials $U_i$ and two-body potentials $V_{ij}$ are considered. All computations are performed in the centre of mass frame where $\bm P = \sum_{i=1}^N \bm p_i$ is vanishing. Natural units, $\hbar = c =1$, will be used throughout the text. 

\subsection{Harmonic oscillators}
\label{sec:oh}

The solutions of Hamiltonian~(\ref{HN}) being approximated within the ET thanks to the solutions of a many-body harmonic oscillator Hamiltonian, it is of prime importance to study the Hamiltonian 
\begin{equation}
\label{Hho}
H_\textrm{ho} = \sum_{i=1}^N \frac{\bm p_i^2}{2 m_i} -\frac{\bm P^2}{2 M} + \sum_{i=1}^N k_i\, \bm s_i^2 + \sum_{i<j=2}^N k_{ij}\, \bm r_{ij}^2
\end{equation}
with $M=\sum_{i=1}^N m_i$. This Hamiltonian can be solved because it can be rewritten as a sum of $N-1$ decoupled harmonic oscillators \cite{silv10}
\begin{equation}
\label{Eohint}
H_\textrm{ho}=\sum^{N-1}_{i=1} \left[ \frac{\bm \sigma_i^2}{2m} + \frac{1}{2} m\, \omega_i^2 \bm z_i^2 \right],
\end{equation}
where $\bm \sigma_i$ and $\bm z_i$ are conjugate variables, and where $m$ is an arbitrary mass scale. The energy of the system is then given by
\begin{equation}
\label{Eohgen}
E_\textrm{ho}=\left\{ 
\begin{aligned}
&\sum^{N-1}_{i=1}  \omega_i\,(2n_i+l_i+D/2) &\quad \textrm{if} \quad D\ge 2 \\
&\sum^{N-1}_{i=1}  \omega_i\,(n_i+1/2) &\quad \textrm{if} \quad D=1
\end{aligned}
\right. .
\end{equation}
where $n_i$ (and $l_i$) are the usual quantum numbers associated with the coordinate $\bm z_i$. These eigenvalues are rid of the centre of mass motion and correspond to states with $\bm P = \bm 0$. The numbers $m\,\omega_i^2/2$ are the eigenvalues of a matrix whose elements are given in \cite{silv10}. Let us note that some parameters $k_i$ or $k_{ij}$ can be negative, provided all values found for $\omega_i^2$ are positive. In some particular situations, these numbers can be analytically obtained. For example, when all particles are identical ($m_i=m$, $k_i=k$ and $k_{ij}=\bar k$, $\forall\ i, j$), then \cite{silv10}
\begin{equation}
\label{Eohid}
E_{{\rm ho}}= \sqrt{\frac{2}{m}(k+N\,\bar k)}\,Q(N) 
\end{equation}
with
\begin{equation}
\label{Qid}
Q(N)=\left\{ 
\begin{aligned}
&\sum^{N-1}_{i=1}  (2n_i+l_i+D/2) &\quad \textrm{if} \quad D\ge 2 \\
&\sum^{N-1}_{i=1}  (n_i+1/2) &\quad \textrm{if} \quad D=1
\end{aligned}
\right. .
\end{equation}
Among other analytical solutions, there is also the case of 3 different particles \cite{silv10}, or when only two types of particles are present (see Sect.~\ref{sec:pract} and \ref{sec:np1}). When the system contains different sets of identical particles, the solution~(\ref{Eohgen}) can be obtained by a procedure which is described in Sect.~\ref{sec:pract}. 

\subsection{Main equations}
\label{sec:main}

The ET is described in \cite{sema18b} for identical particles. The case of different particles is a bit more complex. So, the method is presented here in some detail. The first step to obtain information about $H$ in (\ref{HN}) is to build the auxiliary Hamiltonian $\tilde H$ given by
\begin{align}
\label{Htilde}
\tilde H = &\sum_{i=1}^N\left[ \frac{\bm p_i^2}{2 \mu_i} + T_i(G_i(\mu_i)) -\frac{G_i^2(\mu_i)}{2\mu_i} \right]\nonumber \\
&+ \sum_{i=1}^N \left[ \nu_i\, \bm s_i^2 + U_i(I_i(\nu_i)) - \nu_i\,I_i^2(\nu_i) \right] 
+ \sum_{i<j=2}^N \left[ \rho_{ij}\, \bm r_{ij}^2 + V_{ij}(J_{ij}(\rho_{ij})) - \rho_{ij}\,J_{ij}^2(\rho_{ij}) \right],
\end{align}
where $\mu_i$, $\nu_i$ and $\rho_{ij}$ are parameters to be determined, and $G_i$, $I_i$ and $J_{ij}$ are functions such that  
\begin{subequations}
\label{GIJ} 
\begin{equation}
\label{GIJG} 
T_i'(G_i(x)) - \frac{G_i(x)}{x} = 0,
\end{equation}
\begin{equation}
\label{GIJI}
U_i'(I_i(x)) - 2\,x\,I_i(x)=0,
\end{equation}
\begin{equation}
\label{GIJJ}
V_{ij}'(J_{ij}(x)) - 2\,x\,J_{ij}(x)=0,
\end{equation}
\end{subequations}
where $A'(y)=dA/dy$. It is interesting to put the Hamiltonian~(\ref{Htilde}) under the following form
\begin{equation}
\label{HNtilde}
\tilde H = \sum_{i=1}^N \tilde T_i(p_i) + \sum_{i=1}^N \tilde U_i(s_i) + \sum_{i<j=2}^N \tilde V_{ij}(r_{ij}).
\end{equation}
From (\ref{Htilde}), one can easily obtain the definitions of the auxiliary parts ($\tilde T_i$, $\tilde U_i$, $\tilde V_{ij}$) of the Hamiltonian~(\ref{HNtilde}). Another form is 
\begin{equation}
\label{HNtilde2}
\tilde H = H_\textrm{ho}(\{\mu_{i} \},\{ \nu_{i} \},\{ \rho_{ij} \})+B(\{\mu_{i} \},\{ \nu_{i} \},\{ \rho_{ij} \}),
\end{equation}
where the function $B$ is obtained by subtracting the harmonic oscillator contributions from (\ref{Htilde})
\begin{align}
\label{B}
B(\{\mu_{i} \},\{ \nu_{i} \},\{ \rho_{ij} \}) = &\sum_{i=1}^N\left[ T_i(G_i(\mu_i)) -\frac{G_i^2(\mu_i)}{2\mu_i} \right]\nonumber \\
&+ \sum_{i=1}^N \left[ U_i(I_i(\nu_i)) - \nu_i\,I_i^2(\nu_i) \right] 
+ \sum_{i<j=2}^N \left[ V_{ij}(J_{ij}(\rho_{ij})) - \rho_{ij}\,J_{ij}^2(\rho_{ij}) \right].
\end{align}
An eigenvalue $\tilde E$ of Hamiltonian~(\ref{HNtilde2}) is given by
\begin{equation}
\label{ENtilde2}
\tilde E = E_\textrm{ho}(\{\mu_{i} \},\{ \nu_{i} \},\{ \rho_{ij} \})+B(\{\mu_{i} \},\{ \nu_{i} \},\{ \rho_{ij} \}),
\end{equation}
where $E_\textrm{ho}$ is an eigenvalue of $H_\textrm{ho}$. In this section, it is always assumed that the eigenstates of $\tilde H$, that is to say of $H_\textrm{ho}$, are rid of the centre of mass motion ($\bm P =\bm 0$), as required for the eigensolutions of $H$.

An eigenvalue $\tilde E$ depends on the parameters $\left\{\{\mu_{i} \},\{ \nu_{i} \},\{ \rho_{ij} \}\right\}$. The principle of the method is to search for the set of parameters
\begin{equation}
\label{alpha0}
\alpha_0 = \left\{\{\mu_{i0} \},\{ \nu_{i0} \},\{ \rho_{ij0} \} \right\},
\end{equation}
such that 
\begin{equation}
\label{extrem}
\left. \frac{\partial \tilde E}{\partial \mu_{i}}\right|_{\alpha_0} = 
\left. \frac{\partial \tilde E}{\partial \nu_{i}}\right|_{\alpha_0} =
\left. \frac{\partial \tilde E}{\partial \rho_{ij}}\right|_{\alpha_0} = 0,  \quad \forall\ i,j.
\end{equation}
These conditions can correspond to an extremum or a saddle point. A set $\alpha_0$ depends on the particular state considered. Let us define the Hamiltonian $\tilde H_0$ by
\begin{equation}
\label{Htilde0}
\tilde H_0 = \tilde H(\alpha_0) = H_\textrm{ho}(\alpha_0)+B(\alpha_0).
\end{equation}
If $|\alpha_0\rangle$ is an eigenstate (the quantum numbers are not indicated to lighten the notation) of this Hamiltonian with the eigenvalues $\tilde E_0$, then
\begin{equation}
\label{findG}
\tilde E_0 = \langle \tilde H_0 \rangle_{\alpha_0},
\end{equation}
where $\langle \cdot \rangle_{\alpha_0} = \langle \alpha_0 | \cdot | \alpha_0 \rangle$. According to the reasoning presented in the introduction and detailed in \cite{sema08}, $\tilde E_0$ is then an approximate eigenvalue of the Hamiltonian~(\ref{HN}). 

The application of the Hellmann-Feynman theorem for the parameter $\mu_i$ gives
\begin{align}
\label{Etilde0}
0 = \left. \frac{\partial \tilde E}{\partial \mu_{i}} \right|_{\alpha_0} &
= \left\langle \frac{\partial \tilde H_0}{\partial \mu_{i0}}\right\rangle_{\alpha_0}\nonumber \\
& = \left\langle \frac{G_i^2(\mu_{i0})-\bm p_i^2}{2 \mu_{i0}^2} + G'_i(\mu_{i0}) \left[ T'_i(G_i(\mu_{i0}))-\frac{G_i(\mu_{i0})}{\mu_{i0}} \right] \right\rangle_{\alpha_0}\nonumber \\
& = \frac{1}{2 \mu_{i0}^2} \left( G_i^2(\mu_{i0}) - \langle \bm p_i^2 \rangle_{\alpha_0} \right),
\end{align}
thanks to (\ref{GIJG}). With similar calculations for the parameters $\nu_i$ and $\rho_{ij}$, using relations~(\ref{GIJI}) and (\ref{GIJJ}), one finally obtains
\begin{subequations}
\label{psr0}
\begin{equation}
\label{pi0}
G_i^2(\mu_{i0}) = \langle \bm p_i^2 \rangle_{\alpha_0} = p_{i0}^2,
\end{equation}
\begin{equation}
\label{si0}
I_i^2(\nu_{i0}) = \langle \bm s_i^2 \rangle_{\alpha_0} = s_{i0}^2, 
\end{equation}
\begin{equation}
\label{rij0}
J_{ij}^2(\rho_{ij0}) = \langle \bm r_{ij}^2 \rangle_{\alpha_0} = r_{ij0}^2.
\end{equation}
\end{subequations}
These relations show that the quantities $p_{i0}$,  $s_{i0}$, and $r_{ij0}$ are defined such that $p_{i0}$ is the mean modulus of the momentum for the $i$th particle, $s_{i0}$ the mean distance between this particle and the centre of mass, and $r_{ij0}$ the mean distance between this particle and the $j$th particle. All these observables are directly obtained by the determination of the set $\alpha_0$, and therefore depend on the quantum numbers of the state considered. Using definitions~(\ref{psr0}), relations~(\ref{GIJ}) imply that
\begin{subequations}
\label{GIJprime}
\begin{equation}
\label{GIJG0} 
T_i'(p_{i0}) = \frac{p_{i0}}{\mu_{i0}},
\end{equation}
\begin{equation}
\label{GIJI0}
U_i'(s_{i0}) = 2\,\nu_{i0}\,s_{i0},
\end{equation}
\begin{equation}
\label{GIJJ0}
V_{ij}'(r_{ij0}) = 2\,\rho_{ij0}\,r_{ij0}.
\end{equation}
\end{subequations}

Let us compute $\langle \tilde T_i(p_i) \rangle_{\alpha_0}$ for the set $\alpha_0$. The use of (\ref{pi0}) gives
\begin{equation}
\label{Ttilde0}
\langle \tilde T_i(p_i) \rangle_{\alpha_0} = \frac{\langle \bm p_i^2 \rangle_{\alpha_0}}{2 \mu_{i0}} + T_i(G_i(\mu_{i0})) -\frac{G^2_i(\mu_{i0})}{2\mu_{i0}} = T_i(p_{i0}).
\end{equation}
With similar calculations, it is easy to show that 
\begin{equation}
\label{Epsr}
\tilde E_0 = \sum_{i=1}^N T_i(p_{i0}) + \sum_{i=1}^N U_i(s_{i0}) + \sum_{i<j=2}^N V_{ij}(r_{ij0}).
\end{equation}
With the definitions of $p_{i0}$, $s_{i0}$, $r_{ij0}$, and the structure~(\ref{HN}) of the Hamiltonian under study, the interpretation of (\ref{Epsr}) is quite obvious: each part of the Hamiltonian~(\ref{HN}) is evaluated at a mean value of its argument. This gives a direct estimation for the kinetic and potential contributions. 

Another interesting relation can be obtained by using the virial theorem (see Appendix~\ref{sec:vir}) for $\tilde H_0$. Since the constant terms in this Hamiltonian do not contribute, (\ref{virN}) gives
\begin{equation}
\label{Virho}
\sum_{i=1}^N \frac{\langle \bm p_i^2 \rangle_{\alpha_0}}{\mu_{i0}} =  
2 \sum_{i=1}^N \nu_{i0} \,\langle \bm s_i^2 \rangle_{\alpha_0}
+ 2 \sum_{i<j=2}^N \rho_{ij0} \,\langle \bm r_{ij}^2 \rangle_{\alpha_0}
\end{equation}
Taking into account (\ref{GIJprime}), (\ref{Virho}) finally reduces to
\begin{equation}
\label{VirH}
\sum_{i=1}^N p_{i0}\, T'_i(p_{i0}) =  \sum_{i=1}^N s_{i0}\, U'_i(s_{i0}) + \sum_{i<j=2}^N r_{ij0}\, V'_{ij}(r_{ij0})
\end{equation}
This relation is the equivalent of the virial theorem for the ET. The way to carry out calculations in a practical manner is described in Sect.~\ref{sec:pract}.

If two particles numbered $i$ and $j$ are identical, $p_{i0}^2 = \langle \bm p_i^2 \rangle_{\alpha_0} = \langle \bm p_j^2 \rangle_{\alpha_0} = p_{j0}^2$ since the state $|\alpha_0\rangle$ must be (anti)symmetrized for these two particles. Then, with~(\ref{GIJG0}),
\begin{equation}
\label{papab}
\mu_{i0} = \frac{p_{i0}}{T_i'(p_{i0})} = \frac{p_{j0}}{T_j'(p_{j0})} = \mu_{j0},
\end{equation}
because $T_i=T_j$ since the two particles are identical. It is then clear that parameters $\{\mu_{i0} \}$ are all equal for a set of identical particles. If $\{ i, j, k, l \}$ number any set of particles that are identical in the system and $h$ numbers a different one, similar calculations show that $\nu_{i0}=\nu_{j0}$, $\rho_{ih0}=\rho_{jh0}$, and $\rho_{ij0}=\rho_{kl0}$. A lot of parameters are equal when the system contains identical particles, which can drastically reduce the number of equations (\ref{extrem}) to solve.

When all particles are identical ($T_i=T$, $U_i=U$, $V_{ij}=V$), (\ref{psr0}) reduces to the definition of only two parameters, because $2 N \langle \bm s_i^2 \rangle_{\alpha_0} ={(N-1)\langle \bm r_{ij}^2 \rangle_{\alpha_0}}$ \cite{sema15b}. From (\ref{Eohid}), (\ref{Epsr}), and (\ref{VirH}), it is then possible to determine the approximate eigenvalues with a compact and elegant set of three equations with a nice semiclassical interpretation \cite{sema13,sema18b}
\begin{subequations}
\label{compact}
\begin{equation}
\tilde E_0 = N\, T(p_0) + N\, U \left( \frac{d_0}{N} \right) + C^2_N\, V \left( \frac{d_0}{\sqrt{C^2_N}} \right), \\
\end{equation}
\begin{equation}
d_0\, p_0=Q(N),
\end{equation}
\begin{equation}
N\, p_0\, T'(p_0) =  d_0\, U' \left( \frac{d_0}{N} \right) + \sqrt{C^2_N}\, d_0\, V' \left( \frac{d_0}{\sqrt{C^2_N}} \right),
\end{equation}
\end{subequations}
where $d_0 = N \sqrt{\langle \bm s_i^2 \rangle_{\alpha_0}}$ and $C^2_N=N(N-1)/2$ is the number of particle pairs. But no equivalent set of equations seems to be possible to write when different particles are considered, since (\ref{psr0}) involves a lot of definitions in this case. 

\subsection{Special cases}
\label{sec:special}

For nonrelativistic systems, $T_i(p_i) = \bm p_i^2/(2m_i)$. So, (\ref{GIJG}) implies that $\mu_i = m_i$ but leaves the function $G_i$ undefined. This is not a problem since then $\tilde T_i(p_i) = T_i(p_i)$ and the kinetic part vanishes in $B$. The above calculations are still relevant and, (\ref{Epsr}) and (\ref{VirH}) are still valid. But less parameters need to be determined (see Sect.~\ref{sec:pract}). Let us note that (\ref{pi0}) does not exist and there is consequently no direct access to the values $\langle \bm p_i^2 \rangle_{\alpha_0}$. Nevertheless, the mean value of the kinetic energy can be computed by (\ref{Epsr}) or by (\ref{VirH}), once the approximate energy and the optimal parameters $\left\{\{ \nu_{i0} \},\{ \rho_{ij0} \} \right\}$ are determined. The situation is similar if $U_i(s_i) \propto s_i^2$ or $V_{ij}(r_{ij}) \propto r_{ij}^2$. 

\subsection{Possible bounds}
\label{sec:bounds}

According to (\ref{pi0}) and (\ref{GIJG0}), $\tilde T_i(p_i)$ can be written as
\begin{equation}
\label{Tenv}
\tilde T_i(p_i) = T_i(p_{i0}) + \frac{T'_i(p_{i0})}{2 p_{i0}} \left( p_i^2 - p_{i0}^2 \right).
\end{equation}
This shows that 
\begin{equation}
\label{Tenv2}
\tilde T_i(p_{i0}) = T_i(p_{i0}) \quad \textrm{and} \quad \tilde T'_i(p_{i0}) = T'_i(p_{i0}).
\end{equation}
Kinetic parts $\tilde T_i$ are tangent to $T_i$ at $p_{i0}$, at least. As the quantity $p_{i0}$ depends on quantum numbers, all the tangent functions $\tilde T_i$ for all possible quantum numbers form an envelope of the kinetic parts $T_i$. Envelopes can also be obtained with similar reasonings for the potentials $U_i$ and $V_{ij}$. This is actually the procedure first developed to create the ET and the origin of its name \cite{hall80,hall83}. 

The great interest of the envelopes is to allow the examination of the possible variational character of the ET. Let us define a function $b^T_i$ such that $b^T_i(x^2) = T_i(x)$. Then,
\begin{equation}
\label{Tenv3}
\tilde T_i(p_i) = b^T_i(p_{i0}^2) + {b^T_i}'(p_{i0}^2) \left( p_i^2 - p_{i0}^2 \right).
\end{equation}
If $b^T_i(x)$ is a concave (convex) function for positive values of $x$, then $b^T_i(p_i^2) = T_i(p_i)$ is, in each point, less (greater) than its tangent, that is to say $\tilde T_i(p_i)$. In the same way, we can define $b^U_i(x^2) = U_i(x)$ and $b^V_{ij}(x^2) = V_{ij}(x)$. If $b^U_i(x)$ and $b^V_{ij}(x)$ are concave (convex) functions for positive values of $x$, then $U_i(s_i)$ and $V_{ij}(r_{ij})$ are less (greater) than $\tilde U_i(s_i)$ and $\tilde V_{ij}(r_{ij})$ everywhere, respectively. 

If $T_i(x) \le \tilde T_i(x)$ and $U_i(x) \le \tilde U_i(x)$ and $V_{ij}(x) \le \tilde V_{ij}(x)$, $\forall\, i, j$, and for all positive values of $x$, the comparison theorem \cite{sema11} implies that the eigenvalues of Hamiltonian~($\ref{HN}$) are all less than or equal to the corresponding eigenvalues of Hamiltonian~($\ref{Htilde}$), that is to say the approximate energies of the ET are upper bounds of the energies of the genuine Hamiltonian. Lower bounds are obviously obtained if all inequalities are reversed. 

Finally, if $b^T_i(x)$ and $b^U_i(x)$ and $b^V_{ij}(x)$ are all concave (convex) functions, an approximate ET energy is an
upper (lower) bound of the genuine energy. Conditions~(\ref{extrem}) correspond then to the existence of an extremum in the space of parameters $\left\{\{\mu_{i} \},\{ \nu_{i} \},\{ \rho_{ij} \}\right\}$. In the special situations discussed in Sect.~\ref{sec:special}, the second derivative is vanishing for some of these functions ($b(x)\propto x$). The variational character of the method is then solely ruled by the convexity of the other ones. If these functions are not all concave or convex, the variational character of the solutions cannot be guaranteed, and conditions~(\ref{extrem}) correspond to the existence of a saddle point.

\section{Practical calculations}
\label{sec:pract}

\subsection{Genuine method}
\label{sec:genuine}

After these general considerations about the ET, let us look at a practical method to carry out a calculation, that is to say to determine the set $\alpha_0$ for the particular state considered, and the corresponding energy $\tilde E_0$. A first step is the determination of the eigenvalue of $H_\textrm{ho}$. If the system of $N$ oscillators is composed of $S$ sets of $N_\alpha$ identical particles of mass $\mu_\alpha$, which can be bosons or fermions, the Hamiltonian~(\ref{Hho}) can be written as
\begin{align}
\label{HNS} 
H_\textrm{ho} = &\sum_{\alpha=1}^{S} \sum_{i_\alpha=1}^{N_\alpha} \frac{\bm p_{i_\alpha}^2}{2 \mu_\alpha}  - \frac{\bm P^2}{2 M} 
+ \sum_{\alpha=1}^{S} \sum_{i_\alpha=1}^{N_\alpha} \nu_\alpha\, (\bm r_{i_\alpha}-\bm R)^2 \nonumber \\
&+ \sum_{\alpha=1}^{S} \sum_{{i_\alpha}<{j_\alpha}=2}^{N_\alpha} \rho_{\alpha\alpha}\, (\bm r_{i_\alpha}-\bm r_{j_\alpha})^2 +
\sum_{\alpha<\beta=2}^{S} \sum_{{i_\alpha}=1}^{N_\alpha}\sum_{{j_\beta}=1}^{N_\beta} \rho_{\alpha\beta}\, (\bm r_{i_\alpha}-\bm r_{j_\beta})^2,
\end{align} 
where $N=\sum_{\alpha=1}^{S} N_\alpha$. Defining $M_\alpha=N_\alpha\,\mu_\alpha$ with $M=\sum_{\alpha=1}^{S} M_\alpha$, $\bm P_\alpha = \sum_{i_\alpha=1}^{N_\alpha} \bm p_{i_\alpha}$ and $\bm R_\alpha= \frac{1}{N_\alpha} \sum_{i_\alpha=1}^{N_\alpha} \bm r_{i_\alpha}$, it is a simple matter of calculations to show that ($\rho_{\alpha\beta}=\rho_{\beta\alpha}$)
\begin{subequations}
\label{HNS2}
\begin{equation}
\label{Halphacm}
H_\textrm{ho} = \sum_{\alpha=1}^{S}H_\alpha + H_\textrm{cm} \quad \textrm{with}
\end{equation} 
\begin{equation}
H_\alpha = \sum_{i_\alpha=1}^{N_\alpha} \frac{\bm p_{i_\alpha}^2}{2 \mu_\alpha} - \frac{\bm P_\alpha^2}{2 M_\alpha}
+ \sum_{i_\alpha=1}^{N_\alpha} \nu_\alpha\, (\bm r_{i_\alpha}-\bm R_\alpha)^2
+\sum_{{i_\alpha}<{j_\alpha}=2}^{N_\alpha} \frac{1}{N_\alpha} \left[\sum_{\beta=1}^{S} N_\beta\, \rho_{\alpha\beta}\right]\,(\bm r_{i_\alpha}-\bm r_{j_\alpha})^2,
\end{equation}
\begin{equation}
H_\textrm{cm} = \sum_{\alpha=1}^{S} \frac{\bm P_\alpha^2}{2 M_\alpha} - \frac{\bm P^2}{2 M}
+ \sum_{\alpha=1}^{S} N_\alpha\, \nu_\alpha\, (\bm R_\alpha - \bm R)^2
+ \sum_{\alpha<\beta=2}^{S} N_\alpha\, N_\beta\,\rho_{\alpha\beta}\, (\bm R_\alpha - \bm R_\beta)^2,
\end{equation}
\end{subequations}
where $H_\textrm{cm}$ rules the motion of the centres of mass of the various sets. These relations are a simple generalisation of a result obtained in \cite{hall79}. All these Hamiltonians are decoupled. Their eigenvalues can be computed using the method described in \cite{silv10}. In particular, introducing a set of internal coordinates for each set of particles, the eigenvalues $E_\alpha$ of $H_\alpha$ are given by (see (\ref{Eohid}))
\begin{equation}
\label{Ealpha}
E_{\alpha} = \sqrt{\frac{2}{\mu_\alpha}\left( \nu_\alpha + \sum_{\beta=1}^{S} N_\beta\, \rho_{\alpha\beta}\right)}\,Q(N_\alpha).
\end{equation}
If $E_\textrm{cm}$ is an eigenvalue of $H_\textrm{cm}$, 
\begin{equation}
\label{Ealphac}
E_\textrm{ho} = \sum_{\alpha=1}^{S} E_\alpha + E_\textrm{cm}.
\end{equation}
But the computation of $E_\textrm{cm}$ is more involved. For instance, when $S=2$, 
\begin{equation}
\label{Ec}
E_\textrm{cm} = \sqrt{\frac{2}{M_1 M_2 M}\left( N_1 M_2^2 \nu_1 + N_2 M_1^2 \nu_2 + N_1 N_2 M^2 \rho_{12}\right)}\,Q(2).
\end{equation}
It is generally not possible to find a complete analytical solution. It is then necessary to determine numerically the eigenvalues of $H_\textrm{cm}$ with the procedure schematically described in Sect.~\ref{sec:oh} and detailed in  \cite{silv10}. Let us note that if $S=1$, $N_\alpha = N$ and $H_\textrm{cm}$ vanishes. When one or several sets contain only one particle, a special treatment is necessary. An example is given in the following section.  

The wavefunction for a system containing a set of $N_\alpha$ identical particles must be completely (anti)symmetrized for these $N_\alpha$ particles. Taking into account all the possible characteristics (spin, isospin, colour) of the particles, not all quantum numbers $Q(N_\alpha)$ are allowed for the spatial part. The computation of theses values can be technically very complicated, even for small systems \cite{silv85}. In the following, we will focus on the ground state for particles with only the position as a degree of freedom (see Appendix~\ref{sec:gs}).

Knowing $E_\textrm{ho}$ from (\ref{Ealphac}) and $B$  from (\ref{B}), (\ref{ENtilde2}) is determined and it is possible to solve, at least numerically, the system~(\ref{extrem}) to compute the set of optimal parameters $\alpha_0$, once all the quantum numbers have been fixed. Replacing these parameters in (\ref{ENtilde2}) gives directly the approximate ET energy $\tilde E_0$. If this energy is an upper or a lower bound, it is then easier to find it by an extremisation procedure.  

\subsection{Improvement of the method}
\label{sec:improve}

A source of inaccuracy in the ET is the strong degeneracy due to the global quantum number $Q(N)$. It is inherent to the method which is based on the solutions of the many-body harmonic oscillator Hamiltonian, but it must not appear for general Hamiltonians. For systems with all identical bosons, it has been shown that the modification of $Q(N)$, by adapting a proposal made at the origin for the WKB approximation in \cite{loba09}, allows a noticeable improvement of the ET results \cite{sema15a}. The accuracy is even very strongly improved in the case of Coulomb interactions, for instance. The relevance of this modification of $Q(N)$ has been justified in \cite{sema15b}, by using the ET in combination with a generalisation of the dominantly orbital state (DOS) method \cite{olss97} to $N$-body systems. When $D\ge 2$ (the modification is irrelevant for $D=1$), $Q(N)$ is modified into
\begin{equation}
\label{Qphi}
Q_\phi(N)= \sum^{N-1}_{i=1} (\phi\, n_i+l_i) + (N-1) \frac{D+\phi-2}{2}.
\end{equation}
It is shown that  \cite{sema15b}
\begin{equation}
\label{phi}
\phi=\left\{ 
\begin{aligned}
\sqrt{\beta+2} &\quad \textrm{if} \quad T(p) \propto p^2 \\
\sqrt{\beta+1} &\quad \textrm{if} \quad T(p) \propto p
\end{aligned}
\right. ,
\end{equation}
for power law one-body $U(x)$ or two-body $V(x)$ potentials proportional to $\textrm{sgn}(\beta)\,x^\beta$. In the case of harmonic oscillators, $\phi=2$ as expected, and (\ref{Qid}) is recovered. The price to pay for this improvement is the loss of the variational character of the ET when it is present. 

It is not clear that (\ref{Qphi}) and (\ref{phi}) can be directly applied without modifications in the case of systems with different particles. Moreover, the case of fermions surely necessitates a special treatment. Nevertheless, this will be tested for the Hamiltonians studied in Sect.~\ref{sec:np1}, and denoted improved envelope theory (ITE) in the following. If a gain is achieved in accuracy, it will be an indication that it is worth trying to adapt as best as possible the different global quantum numbers appearing in systems with different particles. 

\section{Systems with one particle different from the other ones}
\label{sec:np1}

To test the reliability of the ET, let us consider a system composed of a first set of $N-1$ particles of type $a$ and one particle, the $N$th, of type $b$. In this case, the second set contains only one particle which can be identified with its centre of mass. In order to avoid confusion between the numbering of the particles and the one for the sets, the index $a$ will be used for the properties of the $N_a$ ($ = N-1$) first particles, and the index $b$ for the last one. Since, the second set contains one particle, its internal Hamiltonian, $H_b$, does not exist, and $\bm R_b = \bm r_N$. So, (\ref{Halphacm}) reduces to $H_\textrm{ho} = H_a + H_\textrm{cm}$ which is written as
\begin{equation}  
\label{HhoNp1}
H_\textrm{ho} = \sum_{i=1}^{N_a} \frac{\bm p_i^2}{2 \mu_a}+\frac{\bm p_N^2}{2 \mu_b}  -\frac{\bm P^2}{2 M} + \sum_{i=1}^{N_a} \nu_a\, \bm s_i^2 + \nu_b\, \bm s_N^2 + \sum_{i<j=2}^{N_a} \rho_{aa}\, \bm r_{ij}^2 + \sum_{i=1}^{N_a} \rho_{ab}\, \bm r_{iN}^2.
\end{equation}
with $M=N_a\,\mu_a+\mu_b$. Such a configuration is chosen because it minimises the number of parameters (no $\rho_{bb}$) and $H_\textrm{ho}$ is analytically solvable. Using the results from the previous section, eigenvalues of this Hamiltonian are given by
\begin{equation}
\label{EhoNp1}
E_\textrm{ho} = \sqrt{\frac{2}{\mu_a}(\nu_a+N_a\, \rho_{aa}+ \rho_{ab})}\, Q(N_a) 
+ \sqrt{\frac{2}{\mu_a\,\mu_b\,M}(\mu_b^2\, \nu_a+N_a\, \mu_a^2\, \nu_b+ M^2\, \rho_{ab})}\, Q(2).
\end{equation}
This formula is a simple generalisation of a result obtained in \cite{hall78}. It can also be obtained directly from (\ref{Eohgen}). We will focus on the ground state (see Appendix~\ref{sec:gs}). So, $Q_\textrm{GS}(2)=D/2$, and $Q_\textrm{GS}(N_a)$ is given by (\ref{QBGS}) or (\ref{QFGS}) depending on the nature of the particle $a$. The function $B$ given by (\ref{B}) takes the following form
\begin{align}
\label{BNp1}
B = &N_a \left[ T_a(G_a(\mu_a)) -\frac{G_a^2(\mu_a)}{2\mu_a} \right] + T_b(G_b(\mu_b)) - \frac{G_b^2(\mu_b)}{2\mu_b}\nonumber \\
&+ N_a \left[ U_a(I_a(\nu_a)) - \nu_a\,I_a^2(\nu_a) \right] + U_b(I_b(\nu_b)) - \nu_b\,I_b^2(\nu_b) \nonumber \\
&+ \frac{N_a(N_a-1)}{2} \left[ V_{aa}(J_{aa}(\rho_{aa})) - \rho_{aa}\,J_{aa}^2(\rho_{aa}) \right] + N_a \left[ V_{ab}(J_{ab}(\rho_{ab})) - \rho_{ab}\,J_{ab}^2(\rho_{ab}) \right].
\end{align}
Once functions $G_i$, $I_i$ and $J_{ij}$ are determined by solving (\ref{GIJ}), $\tilde E$ can be calculated, and the system~(\ref{extrem}) solved to compute the optimal values of parameters $\{\mu_a,\mu_b,\nu_a,\nu_b,\rho_{aa},\rho_{ab}\}$. Their replacement in $\tilde E$ gives then the approximation sought $\tilde E_0$. In the three following considered systems, $D=3$, and positions and momentums are dimensionless quantities. 

\subsection{Simple alternative approximation for $N=3$}
\label{sec:HOB0}

In order to test the accuracy of the ET in the case $N=3$, we will compare our results with the ones obtained from a variational method based on an expansion of a trial three-body state in harmonic oscillator bases with two different oscillator lengths \cite{nunb77}. This method, originally developed for nonrelativistic kinematics, works also very well for relativistic kinematics \cite{silv00,sema01}. But it is quite heavy to use if a high accuracy is searched for. Fortunately, with only one basis state, what is called the ``0 quanta approximation", a reasonable upper bound of the bosonic ground state can be computed with a very simple formula. In this paper, the results of the ET and the ITE for $N=3$ are only compared with this 0 quanta approximation, noted here HOB0. 

\subsection{Ultrarelativistic oscillators}
\label{sec:URoh}

The first system considered is composed of $N$ ultrarelativistic particles with a vanishing mass, interacting via harmonic oscillator potentials. The $N$th particle is different since it has a different interaction with the other ones. The Hamiltonian is written as
\begin{equation}
\label{Huoh}
H = \sum_{i=1}^{N} |\bm p_i| + \sum_{i< j=2}^{N-1} \bm r_{ij}^2 + \lambda \sum_{i=1}^{N-1} \bm r_{iN}^2.
\end{equation}
For this system, the potentials are harmonic ones. So, we are in a special case of Sect.~\ref{sec:special}, with no potential part present in the function $B$, and $\rho_{aa}=1$, $\rho_{ab}=\lambda$ in (\ref{EhoNp1}). Moreover, $\nu_a=\nu_b=0$. With $T_i(p)=p$, $G_i(x)=x$ and (\ref{BNp1}) reduces to
\begin{equation}
\label{Buoh}
B = \frac{1}{2} \left( N_a\,\mu_a+\mu_b \right).
\end{equation}
In this case, $\tilde E_0$ is an upper bound. It has been checked that this value is a minimum of the function $\tilde E(\mu_a,\mu_b)$, and that $\mu_a=\mu_b$ when $\lambda = 1$, as expected. 

When $N=2$, the energy for eigenstates with vanishing angular momentum can be expressed in terms of the zeros of the Airy function Ai \cite{sema04}. From the ground state to high radial excitations, the relative error for the ET upper bounds is less than 7\%. For the IET, the relative errors are less than 3\%, and the energies are no longer upper bounds but are all below the exact values. 

No accurate eigenvalues were found in the literature for the Hamiltonian~(\ref{Huoh}) when $N>2$. Results for some bosonic ground state for $N=3$ in Table~\ref{tab:URoh} show that the ET upper bounds are not as good as the ones from HOB0. The relative error between these results is constant and around 6\%. The relative errors between IET and HB0 results are also constant and less than 1\%. But the IET values are systematically below the HB0 values. According to this limited test, it seems that the accuracy is not degrading as $N$ increases.

\begin{table}[htb]
\protect\caption{Bosonic ground state of Hamiltonian~(\ref{Huoh}) with $N=3$ for several values of $\lambda$. The results from the ET and the IET ($\phi=\sqrt{3}$) are compared with the approximations HOB0.}
\label{tab:URoh}
\begin{tabular}{cccc}
\hline
 & $\lambda=0.1$ &$\lambda=1$ &$\lambda=10$ \\
\cline{2-4}
ET & $5.5971$ & $8.1770$ & $15.3516$ \\
HOB0 & $5.2995$ & $7.7423$ & $14.5354$ \\
IET & $5.2587$ & $7.6826$ & $14.4234$ \\
\hline
\end{tabular}
\end{table} 

\subsection{Nonrelativistic three-body systems with a power law potential}
\label{sec:powlaw}

The second system is composed of three particles, the 3rd one having a different mass, interacting via different power law potentials. The Hamiltonian is written as
\begin{equation}
\label{Hpl}
H = \sum_{i=1}^{2} \frac{\bm p_i^2}{2} + \frac{\bm p_3^2}{2 m}
+ \frac{1}{2}\, \textrm{sgn}(\beta) \sum_{i< j=2}^{3} \bm r_{ij}^\beta .
\end{equation}
For this system, the kinematics is nonrelativistic. So, we are in a special case of Sect.~\ref{sec:special}, with no kinetic part present in the function $B$. Moreover, $N_a=2$, $\mu_a=1$, $\mu_b=m$, $M=m+2$, $\nu_a=\nu_b=0$ in (\ref{EhoNp1}). With $V_{ij}(r)=\frac{1}{2}\, \textrm{sgn}(\beta)\, r^\beta$, $J_{ij}(x)=(4\, x/|\beta|)^{1/(\beta-2)}$ and (\ref{BNp1}) reduces to
\begin{equation}
\label{Bpl}
B = \left( \rho_{aa}^{\beta/(\beta-2)} + 2\, \rho_{ab}^{\beta/(\beta-2)} \right) \left[ \frac{1}{2}\,\textrm{sgn}(\beta) \left( \frac{4}{|\beta|} \right)^{\beta/(\beta-2)} - \left( \frac{4}{|\beta|} \right)^{2/(\beta-2)} \right].
\end{equation}
The case $\beta=2$ is a special one since (\ref{Hpl}) is then exactly a harmonic oscillator Hamiltonian of type (\ref{Hho}) (see Sect.~\ref{sec:special}). It has been checked that $\rho_{aa}\to 1/2$, $\rho_{ab}\to 1/2$, and $\tilde E_0$ tends toward the exact result, as expected, when $\beta \to 2$. It is easy to show that $\tilde E_0$ is an upper (lower) bound when $\beta <2$ ($\beta >2$). It has been checked that this value is the correct extremum of the function $\tilde E(\rho_{aa},\rho_{ab})$, and that $\rho_{aa}=\rho_{ab}$ when $m = 1$, as expected. 

Upper bounds from the ET for the bosonic ground state are given in Table~\ref{tab:pl}. The ``exact" results are obtained with an elaborate hyperspherical expansion up to a grand orbital momentum $L=8$ \cite{basd90}. This insures a good convergence of the expansion and a high accuracy of the eigenvalues \cite{rich81}. For $\beta=-1$, the bound is not good. But for $\beta>0$, the approximation is reasonable, with relative errors around several percents. Results from HOB0 are always better. Results from the IET are much better, with relative errors approximately divided by 10, but the variational character is lost. 

\begin{table}[htb]
\protect\caption{Bosonic ground state of Hamiltonian~(\ref{Hpl}) for several values of $m$ and $\beta$. The results from the ET and the IET ($\phi=\sqrt{\beta+2}$) are compared with the ``exact" ones \cite{basd90} and the approximations HOB0.}
\label{tab:pl}
\begin{tabular}{ccccccccccc}
\hline
 & & \multicolumn{4}{c}{$m=0.2$} & &  \multicolumn{4}{c}{$m=5$} \\
\cline{3-6}\cline{8-11}
$\beta$ & \  & ET & IET & ``exact" & HOB0 & & ET & IET &``exact"  & HOB0 \\ 
\hline
$-1$ \ & & $-0.0645$ & $-0.1452$ & $-0.1398$ & $-0.1232$ & \ & $-0.1797$ & $-0.4043$ & $-0.3848$ & $-0.3432$ \\
$0.1$ & & $1.9804$ & $1.9425$ & $1.9452$ & $1.9480$ & & $1.8820$ & $1.8460$ & $1.8486$ & $1.8512$ \\
$1$ & & $5.2278$ & $4.9117$ & $4.9392$ & $4.9498$ & &  $3.6386$ & $3.4186$ & $3.4379$ & $3.4451$ \\
$3$ & & $8.9925$ & $9.8482$ & $9.7389$ & $9.7639$ & &  $4.6320$ & $5.0728$ & $5.0166$ & $5.0293$ \\
\hline
\end{tabular}
\end{table}

\subsection{Atoms}
\label{sec:atoms}

In atomic units, the Hamiltonian for $N_e$ electrons in an atom of charge $Z$ is written as
\begin{equation}
\label{HAt}
H = \frac{1}{2} \sum_{i=1}^{N_e} \bm p_i^2 + \frac{1}{2 m} \bm p_{N}^2 
- Z\sum_{i=1}^{N_e} \frac{1}{|\bm r_{iN}|} + \sum_{i< j=2}^{N_e} \frac{1}{|\bm r_{ij}|},
\end{equation}
where $N=N_e+1$ is the number of the nucleus with a mass $m$. Energies in eV are obtained by multiplying the eigenvalues of $H$ by the usual factor $\alpha^2\, m_e = 27.21$~eV. For this system, the kinematics is also nonrelativistic and no kinetic part is present in the function $B$. Moreover, $N_a=N_e$, $\mu_a=1$, $\mu_b=m$, $M=m+N_e\approx m$ ($m\ge 1836.15$), $\nu_a=\nu_b=0$ in (\ref{EhoNp1}). For $V_{ab}(r)=-Z/r$, $J_{ab}(x)= (\frac{Z}{2x})^{1/3}$ with $x>0$. For $V_{aa}(r)=1/r$, $J_{aa}(x)= (\frac{-1}{2x})^{1/3}$ with $x <0$ because of the repulsive nature of the interaction between two electrons. So, (\ref{BNp1}) reduces to
\begin{equation}
\label{BAt}
B = \frac{3}{2^{2/3}}\left[ \frac{N_e(N_e-1)}{2}\,|\rho_{aa}|^{1/3} - N_e\, Z^{2/3}\,\rho_{ab}^{1/3}  \right].
\end{equation}
Let us note that $\rho_{ab}>0$ and $\rho_{aa}<0$ in (\ref{EhoNp1}) and in (\ref{BAt}). The variational character of $\tilde E_0$ cannot be determined due to the mixing of attractive and repulsive potentials. It has been checked that this value is a saddle point of the function $\tilde E(\rho_{aa},\rho_{ab})$. 

Hamiltonian~(\ref{HAt}) does not contain all the interactions present in an atom (spin effects and relativistic corrections are neglected), but the main contribution is taken into account. So, the results of the ET are compared with the experimental data about ionisation energies \cite{nist} which are certainly very close to the eigenvalues of (\ref{HAt}). Taking into account the fermionic nature of the electrons, ground state binding energies for some atoms computed with the ET are presented in Table~\ref{tab:At} with the experimental values, and the approximations HOB0 when $N=3$. Results from the ET are not good, as it could be expected from the study of the previous system. The results from the IET are only indicated for atoms in which the fermionic nature of the electrons can be ignored, that is to say for $N_e = 1$ and $2$. The improvement is dramatic. The binding energies are exact for $N_e = 1$. For $N_e = 2$, the relative errors are divided by 10 or more, and are smaller than for the HOB0 upper bounds. Some very naive tests have convinced us that a dramatic improvement is also possible for atoms with $N_e \ge 3$. But, to obtain relevant values in this case, it will be necessary to generalise the technique developed in \cite{sema15b} to fermions. 

\begin{table}[htb]
\protect\caption{Ground state binding energies (in eV) of Hamiltonian~(\ref{HAt}) for some atoms. Results from the ET and the IET ($\phi = 1$, $N_e \le 2$) are compared with the experimental values (see text), and the approximations HOB0 when $N=3$.}
\label{tab:At}
\begin{tabular}{ccccc}
\hline
 & ET & IET & Exp. & HOB0 \\
\cline{2-5}
H & 6.0 & 13.6 & 13.6 & - \\
$^4$He$^+$ & 24.2 & 54.4 & 54.4 & - \\
$^4$He & 33.1 &  74.5 & 79.0 & 63.2 \\
$^6$Li$^+$ & 85 & 191 & 198 & 162 \\
$^6$Li & 66 & - & 203 & - \\
$^{12}$C$^{4+}$ & 386 & 868 & 882 & 737 \\
$^{12}$C & 321 & - & 1030 & - \\
$^{16}$O$^{6+}$ & 707 & 1591 & 1611 & 1351 \\
$^{16}$O & 672 & - & 2044 & - \\
\hline
\end{tabular}
\end{table}

\section{Conclusion}
\label{sec:conclu}

In the case of quantum systems with all identical particles, approximate eigenvalues can be computed in the framework of the envelope theory by solving a compact and elegant set of three equations (\ref{compact}) \cite{sema13,sema17,sema18b,sema19}. The accuracy of the eigenvalues can be improved by modifying the global quantum number associated with a state \cite{sema15a}, according to considerations provided by the dominantly orbital state method \cite{sema15b}. The improvement is sometimes marginal, as for Gaussian potentials, or sometimes big, as for Coulomb potentials. The main goal of the (improved) envelope theory is to provide rapidly reliable solutions if a great accuracy is not searched for. The approximate eigenvalues obtained can be used for instance as tests for more accurate numerical calculations. The method is very easy to implement whatever the number of particles, and allows the treatment of non usual kinetic energy. 

In this work, it is shown that the envelope theory can be extended to treat also many-body systems with different particles. General considerations about the method are presented as well a practical manner to compute the eigenvalues. The accuracy is tested with three different systems: ultrarelativistic oscillators, nonrelativistic three-body systems with a power law potential, and atoms. As in the case of identical particles, if fairly good results can be obtained with the genuine envelope theory for some kind of potentials, this is not the case for other ones. Nevertheless, a very crude application of the improvement procedure for identical particles to systems with different particles can provide a dramatic improvement of the accuracy. So, it is worth trying to adapt as best as possible this improvement procedure to many-body systems with different particles. This will be the subject of another publication. In order to perform more tests of the generalisation of the envelope theory presented here, it will be desirable to have available accurate computations of eigenvalues for large many-body systems containing different particles, bosons or fermions, as the ones performed in \cite{horn14} for identical bosons. 

\begin{acknowledgments}
This work was supported by the Fonds de la Recherche Scientifique - FNRS under Grant Number 4.4510.08. 
\end{acknowledgments} 

\appendix

\section{Centre of mass}
\label{sec:cm}

The centre of mass of the system ruled by the Hamiltonian~$\tilde H_0$, which is the best approximant for $H$, is given by
\begin{equation}
\label{cmHtilde0}
\bm R = \frac{\sum_{i=1}^N \mu_{i0} \,\bm r_i}{\sum_{i=1}^N \mu_{i0}}.
\end{equation}
If the kinematics of $H$ is nonrelativistic, then $\mu_{i0} = m_i$ (see Sect.~\ref{sec:special}), and the definition~(\ref{cmHtilde0}) is the usual one. Let us look at the meaning of $\mu_{i0}$ for an arbitrary kinematics $T_i(p_i)$ with $p_i=|\bm p_i|$. Within the Hamiltonian formalism, $\dot{\bm r}_i=\partial H/\partial \bm p_i$, that is to say $\dot{\bm r}_i=\partial T_i/\partial \bm p_i$ for (\ref{HN}). Since $T_i$ depends only on $p_i$, $\dot{\bm r}_i\parallel \bm p_i$ \cite{sema18a} and $|\dot{\bm r}_i|=dT_i(p_i)/dp_i$. Within the ET, $T_i'(p_{i0})$ can then be interpreted as the mean speed $v_{i0}$ of the $i$th particle. From~(\ref{GIJG0}), we have
\begin{equation}
\label{mui0}
p_{i0} = \mu_{i0}\,v_{i0},
\end{equation}
and $\mu_{i0}$ plays the role of an effective mass for this particle.

For instance, $T(p)=p^2/(2 m)$ gives $\mu(p)=p/T'(p)=m$, as expected. If $T(p)=\sqrt{p^2+m^2}$, then $\mu(p)=T(p)$, which is the correct form to compute the centre of mass for a system of free relativistic particles. So, one can consider that formula~(\ref{cmHtilde0}), taking into account only quantities from the kinetic part, yields a reasonable estimation for the centre of mass position in the case of arbitrary kinematics. Indeed, the force fields in (\ref{HN}) carry only energy and no momentum, contrary to the kinetic part which conveys both energy and momentum.

\section{Virial theorem}
\label{sec:vir}

The quantum virial theorem has been generalised for arbitrary kinematics in the case of one/two-body Hamiltonian \cite{luch90}. Using the Hellmann-Feynman theorem as in \cite{ipek16}, it is easy to show, for the $N$-body Hamiltonian
\begin{equation}
\label{Hgen}
H_N = \sum_{i=1}^N T_i(\bm p_i) + \sum_{i=1}^N U_i(\bm s_i) + \sum_{i<j=2}^N V_{ij}(\bm r_{ij}),
\end{equation}
that
\begin{equation}
\label{virN}
\sum_{i=1}^N  \left\langle \bm p_i\cdot \frac{\partial T_i(\bm p_i)}{\partial \bm p_i}\right\rangle 
= \sum_{i=1}^N \left\langle \bm s_i\cdot \frac{\partial U_i(\bm s_i)}{\partial \bm s_i} \right\rangle 
+ \sum_{i<j=2}^N \left\langle \bm r_{ij}\cdot \frac{\partial V_{ij}(\bm r_{ij})}{\partial \bm r_{ij}}\right\rangle,
\end{equation}
where the mean value is computed with a stationary eigenstate of $H_N$. 

\section{Ground state}
\label{sec:gs}

An approximate eigenstate of the Hamiltonian~(\ref{HN}) is an eigenstate of the Hamiltonian~(\ref{Htilde0}). So the symmetry and other characteristics of the approximate solutions are given by the properties of the solutions of the many-body harmonic oscillator Hamiltonian~(\ref{Hho}). Besides position and momentum, particles can be characterised by a lot of degrees of freedom, such as spin, isospin, color. We will here focus only on the spatial symmetry of the many-body states. If all particles are different, then all quantum numbers are allowed and the ground state is achieved when all these numbers are vanishing. If the system contains several sets of identical particles which can be bosons or fermions, it is useful to consider the alternative form~(\ref{HNS2}). The part $H_\textrm{cm}$ rules the motion of the centres of mass of the different sets, so no symmetry must be considered and the ground state is achieved when all the corresponding quantum numbers are vanishing. The ground state of each Hamiltonian $H_\alpha$ can be computed using the procedure developed in \cite{levy68}. It is shown that the internal energy of a system of $N_\alpha$ identical particles, interacting via two-body harmonic forces, is equivalent to the energy of a system of uncoupled oscillators in a common central field, provided the contribution of the centre of mass is removed. This result can be directly extended if one-body harmonic forces are added. So the ground state of theses particles can be computed as the ground state of $N_\alpha-1$ independent particles. 

The bosonic ground state is trivial to compute and is given by 
\begin{equation}
\label{QBGS}
Q_\textrm{BGS}(N_\alpha) = (N_\alpha-1)\frac{D}{2},
\end{equation}
with all quantum numbers vanishing in (\ref{Qid}). The corresponding eigenstate is the product of Gaussian functions and is completely symmetric. 

For fermions, with degeneracy $d$, the situation is more complicated. In this case, the level of a harmonic oscillator with an energy $q$ can accept 
\begin{equation}
\label{levelq}
d \begin{pmatrix}
q + D-1 \\ D-1
\end{pmatrix} 
\end{equation}
particles. By piling particles on levels with higher energies and taking into account that the last level can be partly occupied \cite{levy68}, the fermionic ground state can be computed with the two following relations   
\begin{subequations}
\label{QFGS}
\begin{equation}
\label{QFGS1}
Q_\textrm{FGS}(N_\alpha) = d\, D \begin{pmatrix}
q + D-1 \\ D+1
\end{pmatrix} + q\, r + (N_\alpha-1)\frac{D}{2} \quad \textrm{with}
\end{equation}
\begin{equation}
\label{QFGS2}
N_\alpha = d \begin{pmatrix}
q + D-1 \\ D
\end{pmatrix} + r,
\end{equation}
\end{subequations}
where $q$ is the greatest natural number such that $r\ge 0$ in (\ref{QFGS2}). If $N_\alpha \gg 1$, an approximate formula is
\begin{equation}
\label{QFGSapp}
Q_\textrm{FGS}(N_\alpha) \approx \frac{D}{D+1}\left(\frac{D!}{d}\right)^{1/D}N_\alpha^{(D+1)/D}.
\end{equation}

\end{document}